\documentclass[prd,twocolumn,floatfix,superscriptaddress,nofootinbib]{revtex4}
\usepackage[utf8]{inputenc}
\usepackage{graphicx,epsfig}
\usepackage{graphicx}
\usepackage{amsmath}
\usepackage{amssymb}
\usepackage{multirow}
\usepackage{float}
\usepackage[usenames,dvipsnames]{color}
\usepackage[colorlinks]{hyperref}

\hypersetup{
     breaklinks=true,
     pdfstartview={FitH},
     colorlinks=true,
     linkcolor=blue,
     citecolor=red,
     filecolor=magenta,
     urlcolor=blue,
     anchorcolor=green,
     linktocpage=true
}

\begin{document}



\begin{center}
\large \bf {Nonlinear 
Scalarization of Reissner-Nordström Black Holes
 in Scalar-Torsion-Maxwell Teleparallel Gravity}
\end{center}

\author{Joaqu\'in Robledo}
\email{joaquin.montenegro@userena.cl}
\affiliation{Departamento de F\'isica, Facultad de Ciencias, Universidad de La Serena,\\
Avenida Cisternas 1200, La Serena, Chile.}
\affiliation{Departamento de F\'isica, Facultad de Ciencias, Universidad de Tarapac\'a,\\
Casilla 7-D, Arica, Chile.}
\affiliation{Instituto de F\'isica y Astronom\'ia, Universidad de Valpara\'iso,\\
Avenida Gran Breta\~na 1111, Valpara\'iso, Chile.}

\author{Yerko V\'asquez}
\email{yvasquez@userena.cl}
\affiliation{Departamento de F\'isica, Facultad de Ciencias, Universidad de La Serena,\\
Avenida Cisternas 1200, La Serena, Chile.}

\author{P. A. Gonz\'{a}lez}
\email{pablo.gonzalez@udp.cl} 
\affiliation{Facultad de Ingenier\'{i}a y Ciencias, Universidad Diego Portales, Avenida Ej\'{e}rcito Libertador 441, Casilla 298-V, Santiago, Chile.}


\begin{abstract}


We investigate the nonlinear scalarization of Reissner–Nordström black holes in scalar-torsion-Maxwell teleparallel gravity, in which the scalar field is nonminimally coupled to the Weitzenböck invariant. We consider polynomial coupling functions that avoid tachyonic instabilities at the linear level, thereby exploring a scalarization mechanism driven by nonlinear effects. In the decoupling limit, we study the nonlinear evolution of finite-amplitude scalar perturbations and show that the inclusion of a sixth-order coupling term allows the scalar field to dynamically evolve toward nontrivial stationary configurations. This behavior can be understood in terms of the effective potential, which develops nontrivial local minima that provide a mechanism for nonlinear saturation. By numerically solving the fully coupled field equations, we construct asymptotically flat charged black hole solutions endowed with scalar hair. We classify these solutions according to the number of nodes in the scalar field profile, identifying multiple branches whose structure and multiplicity depend on the coupling parameters. We further show that the higher-order coupling progressively suppresses higher-node branches and characterize the scalar backreaction through deviations of the metric functions and the torsion scalar from the Reissner–Nordström background.


\end{abstract}

\maketitle



\clearpage


\section{Introduction}

According to the no-hair theorem \cite{Wheeler, Israel, Carter, Hawking, Robinson}, stationary, asymptotically flat black holes in General Relativity (GR) are uniquely characterized by their mass $M$, electric charge $Q$, and angular momentum $J$. Nevertheless, viable extensions of GR, such as scalar-tensor theories \cite{Fujii}, have been extensively investigated to determine whether black holes can evade this uniqueness result and develop additional degrees of freedom in the form of scalar hair.

When a scalar field is nonminimally coupled to gravity, its backreaction on the spacetime geometry can give rise to black hole solutions endowed with scalar hair. One of the earliest counterexamples to the no-hair theorem was the Bocharova-Bronnikov-Melnikov-Bekenstein (BBMB) solution \cite{1970MVSFA..11..706B,Bekenstein:1974sf}, which describes a black hole with a scalar field conformally coupled to gravity. An extension of this solution including an electromagnetic field was subsequently presented in Ref. \cite{Bekenstein:1975ts}. Although this configuration possesses ``secondary hair'', it is physically problematic because the scalar field diverges at the event horizon, and the solution exhibits  stability issues  \cite{bronnikov}. However, this divergence can be avoided by introducing a cosmological constant, which allows for regular scalar field configurations at the event horizon. Such solutions have been investigated in various gravitational theories and in different asymptotic backgrounds, including asymptotically anti-de Sitter (AdS) and de Sitter (dS) spacetimes \cite{Martinez:1996gn, Martinez:2002ru,  Winstanley:2002jt, Kolyvaris:2010yyf, Gonzalez:2013aca, Gonzalez:2014tga, Charmousis:2014zaa, Barrientos:2016ubi, Herrera:2017ztd, Filios:2018xvy}.

Beyond these solutions, black holes can also develop scalar hair dynamically through a mechanism known as ``scalarization''. This phenomenon occurs when strong gravitational effects trigger an instability of the scalar free black hole solution, allowing the scalar field to grow dynamically and develop a nontrivial profile around the event horizon.

Originally proposed by Damour and Esposito-Farèse \cite{Damour:1993hw}, spontaneous scalarization was first studied in the context of neutron stars, where a tachyonic instability can arise from a nonminimal coupling between the scalar field and matter \cite{Berti:2015itd}. This mechanism allows scalar-tensor theories to satisfy weak-field constraints from Solar System experiments while predicting potentially large deviations from GR in the strong-field regime. Such deviations can produce distinctive observational signatures in binary pulsar systems \cite{Dcited2} and compact object mergers \cite{Ramazanoglu:2016kul}.

The scalarization mechanism has been extended to black hole spacetimes by replacing the matter coupling with a non-minimal coupling to curvature or gauge invariants. For example, by introducing a non-minimal coupling between a scalar field and the electromagnetic Maxwell invariant, the standard Reissner-Nordström (RN) black hole becomes unstable against scalar perturbations \cite{Herdeiro:2018wub}. This mechanism has motivated several extensions of the scalarization framework. For instance, the scalarization of the RN black hole in the presence of higher-derivative gauge field corrections was recently investigated in \cite{Kiorpelidi:2023jjw}. Furthermore, this phenomenon has even been studied in the context of gauge/gravity duality, where the holographic scalarization of a charged scalar field was found to occur below a critical phase-transition temperature \cite{Guo:2020sdu}.

Beyond spontaneous scalarization, the formation and dynamics of scalarized black holes in scalar-Gauss-Bonnet gravity have been extensively studied \cite{Ripley:2020vpk,Silva:2020omi,Doneva:2021dqn,Kuan:2021lol,East:2021bqk, Zou:2024wsk}. In Ref. \cite{Doneva:2021tvn}, the authors demonstrated the existence of a fully nonlinear dynamical mechanism for the formation of scalarized black holes that differs from standard spontaneous scalarization, as it does not rely on a linear tachyonic instability. Although Schwarzschild black holes are linearly stable against scalar perturbations in this context, certain choices of the coupling function render them unstable against nonlinear scalar perturbations. This nonlinear instability can ultimately lead to the formation of new black hole solutions endowed with scalar hair.  In addition to these developments, the dynamical evolution of scalar hair has been investigated in the context of binary black hole coalescence in Einstein–Maxwell–scalar theory \cite{Diaz:2026iyx}.

While the scalarization mechanisms discussed above have primarily been explored in curvature-based theories of gravity, scalarization can also arise within Teleparallel Gravity (TG), a framework in which the gravitational interaction is formulated in terms of spacetime torsion rather than curvature. TG has been extensively studied and extended in several directions, leading to a variety of generalized theories, including Teleparallel Supergravity \cite{Salgado:2005pg}, scalar-torsion theories \cite{Geng:2011aj, Gonzalez:2014pwa, Kofinas:2015hla, Kofinas:2015zaa, Hohmann:2018vle, Hohmann:2018dqh, Hohmann:2018ijr}, Kaluza-Klein reductions \cite{Geng:2014nfa}, the Teleparallel Equivalent of Gauss-Bonnet Gravity \cite{Kofinas:2014owa}, the Teleparallel Equivalent of Lovelock Gravity \cite{Gonzalez:2015sha,Gonzalez:2019tky}, higher-dimensional teleparallel theories \cite{Astudillo-Neira:2017anx}, and novel classes of modified TG models \cite{Bahamonde:2017wwk}.

Building on these developments, black hole scalarization has also been explored within the teleparallel framework. Early contributions in this direction included the construction of exact three-dimensional hairy black hole solutions in scalar-torsion gravity \cite{Gonzalez:2014pwa}. In four-dimensional asymptotically flat spacetimes, torsion-induced scalarization by the coupling to the teleparallel Gauss-Bonnet invariant has been shown to produce black holes with novel nonmonotonic scalar field profiles, distinguishing them from their standard curvature-based counterparts \cite{Bahamonde:2022chq, Bahamonde:2021srr, Bahamonde:2022lvh}. More recently, the fully nonlinear scalarization of Schwarzschild black holes in scalar-torsion teleparallel gravity was investigated in \cite{Gonzalez:2024ifp}. Extending this framework to electrically charged configurations provides a natural setting in which to investigate how higher-order scalar-torsion couplings affect the existence and multiplicity of nonlinear scalarized black hole branches. Furthermore, scalar-torsion dynamics have recently been applied to cosmological scales, including the production of primordial black holes during inflation in $f(T)$ gravity \cite{Villalobos-Silva:2026deu}.

In this work, we investigate the nonlinear scalarization of RN black holes in scalar-torsion-Maxwell teleparallel gravity, employing polynomial coupling functions that avoid tachyonic instabilities at the linear level. We first examine the nonlinear evolution of finite-amplitude scalar perturbations in the decoupling limit. We show that, while the purely quartic coupling leads to a nonlinear runaway instability for sufficiently large perturbations, the inclusion of a sixth-order coupling term allows the scalar field to dynamically evolve toward nontrivial stationary configurations. This behavior is physically understood through the effective potential, which develops nontrivial local minima that provide a mechanism for the nonlinear saturation of scalar perturbations. By numerically solving the fully coupled field equations, we then construct asymptotically flat charged black hole solutions endowed with scalar hair. We classify these solutions according to the number of nodes in the scalar field profile, identifying multiple branches whose structure and multiplicity depend on the coupling parameters. In particular, we show that the higher-order coupling modifies the nodal branch structure and progressively suppresses higher-node solutions as its strength increases. Finally, we investigate the backreaction of the scalar field on the spacetime geometry by analyzing the deviations of the metric functions and the torsion scalar from the RN background.


The work is organized as follows: In Sec. \ref{MGT}, we provide a brief review of the Teleparallel Equivalent of General Relativity. In Sec. \ref{Models}, we present the Scalar-Torsion-Maxwell model and derive the equations of motion. In Sec. \ref{decouplinglimit}, we analyze the dynamics of the scalar field in the decoupling limit over a fixed RN background. We then numerically construct and study the fully coupled scalarized RN black holes in Sec. \ref{numeric}. Finally, our conclusions are presented in Sec. \ref{conclusion}.

\section{Teleparallel Equivalent of General Relativity}
\label{MGT}

The Teleparallel Equivalent of General Relativity (TEGR) \cite{Unzicker:2005in,TEGR1,TEGR2,Aldrovandi:2013wha,Arcos:2004tzt,Maluf:2013gaa, Pereira:2013qza}  is an alternative formulation of GR in which the gravitational interaction is described in terms of spacetime torsion rather than curvature.  Unlike the standard formulation of GR, which is based on the Levi-Civita connection, TEGR employs the Weitzenb\"{o}ck connection to define the covariant derivative. This connection is curvature-free but possesses non-vanishing torsion, so that the gravitational degrees of freedom are encoded in the torsion tensor. Despite this different geometrical description, TEGR is dynamically equivalent to GR. This equivalence follows from the relation between the Ricci scalar associated with the Levi-Civita connection and the torsion scalar constructed from the Weitzenb\"{o}ck connection, namely,
\begin{equation}
\mathcal{T}\, =\, -R + 2\; e^{-1}\;\partial _{\nu }(e\,T_{\sigma }^{\ \sigma \nu
}\,)\,,  \label{divergence}
\end{equation}
where $R$ denotes the Ricci scalar and $e$ is the determinant of the vielbein. The second term on the right-hand side is a total divergence, which ensures that the TEGR and the Einstein-Hilbert actions differ only by a boundary contribution. Consequently, both formulations yield equivalent field equations under the same boundary conditions. The quantity $\mathcal{T}$ in Eq.~(\ref{divergence}) is the Weitzenb\"{o}ck invariant, also known as the torsion scalar, and is defined as 
\begin{equation} \label{scalar-T}
\mathcal{T}\ =\ S_{\rho }^{\ \mu \nu }\ T^{\rho }_{\ \mu \nu }\ =\  \frac{1}{4}T^{\rho \mu \nu}T_{\rho \mu \nu} + \frac{1}{2}T^{\rho \mu \nu}T_{ \nu \mu \rho} - T_{\rho \mu}{}^{\rho}T^{\nu \mu}{}_{\nu}\,,
\end{equation}
where $T^{\rho }_{\ \mu \nu }$ denotes the torsion tensor. In the Weitzenb\"{o}ck gauge, where the spin connection vanishes, the torsion two-form is simply
$T^{a}=de^{a}$. In a coordinate basis, the corresponding Weitzenb\"{o}ck connection is given by $\Gamma^\lambda_{\nu%
\mu}=\,e_{a}^{\lambda }\,\partial _{\nu }e_{\mu }^{a}$, where $e_{\mu }^{a}$ and $e_{a}^{\mu}$ are the vielbein and its inverse, respectively. The superpotential $S_{\rho}^{\,\,\, \mu \nu}$, which is antisymmetric in its last two indices, is defined as
\begin{equation}
S_{\ \mu \nu }^{\rho }=\frac{1}{4}\,(T_{\ \mu \nu }^{\rho }-T_{\mu \nu }^{\
\ \ \rho }+T_{\nu \mu }^{\ \ \ \rho })+\frac{1}{2}\ \delta _{\mu }^{\rho }\
T_{\sigma \nu }^{\ \ \ \sigma }-\frac{1}{2}\ \delta _{\nu }^{\rho }\
T_{\sigma \mu }^{\ \ \,\sigma }.  \label{S}
\end{equation}
In $d$ spacetime dimensions, the fundamental dynamical variables are the $d$ linearly independent vielbein fields $e_{\mu }^{a}$. The torsion tensor is constructed from these fields and their first derivatives. In particular, the torsion scalar $\mathcal{T}$ is a specific quadratic combination of the irreducible components of the torsion tensor under the Lorentz group $SO(1,3)$ \cite{Hehl:1994ue}. 
In TEGR, the torsion tensor 
contains the information necessary to describe the
gravitational field, and the action is given by
\begin{eqnarray}  \label{action0}
I = \frac{1}{2 \kappa}\int d^4x e \left(\mathcal{T}+\mathcal{L}_{m}\right),
\end{eqnarray}
where $\kappa =8 \pi G$, $e = \text{det}(e_{\,\, \mu}^a) = \sqrt{-g}$, $\mathcal{T}$ is the torsion scalar, and $\mathcal{L}_{m}$ stands for the matter Lagrangian. 
The equations of motion can be obtained through the variation of the action (\ref{action0}) with respect to the vielbein, which yields
\begin{eqnarray}\label{eom}
e^{-1}\partial_{\mu}(eS_{a}{}^{\mu\nu})
-e_{a}^{\lambda}T^{\rho}{}_{\mu\lambda}S_{\rho}{}^{\nu\mu}
-\frac{1}{4}e_{a}^{\nu
}\mathcal{T}
= 4\pi Ge_{a}^{\rho}\overset {\mathbf{em}}T_{\rho}{}^{\nu}\,,
\end{eqnarray}
where we have introduced the mixed-index superpotential $S_a{}^{\mu\nu} =
e_a^{\,\,\, \rho}S_{\rho}{}^{\mu\nu}$ and
$\overset{\mathbf{em}}{T%
}_{\rho}{}^{\nu}$ denotes the usual energy-momentum
tensor of the matter fields. The vielbein fields ${\mathbf{e}_a(x^\mu)}$, form an orthonormal
basis for the tangent space at each point $x^\mu$ of the manifold, that is
$\mathbf{e} _a\cdot\mathbf{e}_b=\eta_{ab}$, with
$\eta_{ab}=diag (1,-1,-1,-1)$. Moreover, the vector $\mathbf{e}_a$ can
be expressed in terms of its components $e_a^{\,\,\, \mu}$ in a coordinate basis,
namely $\mathbf{e}_a=e_a^{\,\,\, \mu}\partial_\mu$. Thus, although TEGR describes gravitation through spacetime torsion rather than curvature, its field equations are dynamically equivalent to those of GR. This equivalence provides the basis for extending the theory to modified teleparallel models, in which the gravitational dynamics can be generalized by introducing nonlinear functions of the torsion scalar or nonminimal couplings to additional fields, as considered in the present work.

\section{Model of Nonlinear Scalarization}
\label{Models}

In this work, we consider the scalar-torsion-Maxwell (stM) model, described by the action 
\begin{equation}
\resizebox{\columnwidth}{!}{$\displaystyle
    I =\int d^{4}x\,e \left[ \left(\frac{1}{2 \kappa}+ f(\phi) \right)\mathcal{T}+\frac{1}{2} g^{\mu \nu} \partial_{\mu} \phi \partial_{\nu}\phi
    - \frac{1}{2\kappa}F_{\mu \nu } F^{\mu \nu }\right] \,,$}
\label{L0}
\end{equation}
where $\kappa=8\pi G$, $\phi$ is the scalar field and $f(\phi)$ is the coupling function of the scalar field to the Weitzenb\"{o}ck invariant. The Maxwell tensor is defined as $F_{\mu\nu}=\partial_{\mu}A_{\nu}-\partial_{\nu}A_{\mu}$, where $A_{\mu}$ is the electromagnetic four-potential. For the static, spherically symmetric configurations considered here, we adopt the ansatz $A_{\mu}dx^{\mu}=\Psi(r)dt$, where $\Psi (r)$ denotes the electrostatic potential.  

We aim to investigate nonlinear scalarized black-hole solutions \cite{Doneva:2021tvn, Staykov:2022uwq} in the stM model (\ref{L0}); therefore, we consider coupling functions that meet the following conditions
\begin{equation}\label{conditions}
\frac{df}{d\phi}(0)=0\,,\quad \frac{d^2 f}{d\phi^2}(0)=0\,.    
\end{equation}
The first condition guarantees that the scalar field equation admits the trivial configuration $\phi = 0$, so that the RN solution is also a solution of the field equations. The second condition eliminates the contribution of the nonminimal coupling to the effective mass of linear scalar perturbations around the RN background, i.e., $m^2_{\text{eff}}=0$. Consequently, the RN solution does not develop a tachyonic instability in the linear scalar sector. Therefore, any scalarized solutions that arise under these conditions cannot be attributed to the standard tachyonic mechanism of spontaneous scalarization. Instead, they may emerge through nonlinear effects, giving rise to scalarized black hole branches that coexist with the RN branch.

The polynomial coupling functions considered in this work, which satisfy the conditions given in Eq. (\ref{conditions}), are
\begin{equation}
    f_1(\phi)=\eta \phi^4 \,, \quad f_2(\phi)=\eta \phi^4 - \gamma\phi^6\,,
\end{equation}
where $\eta$ and $\gamma$ are coupling parameters.

Varying the action with respect to the vierbein, the scalar field, and the electromagnetic four-potential yields the corresponding equations of motion
\begin{eqnarray}\label{eomst}
\notag    && \left(\frac{2}{\kappa}+4 f(\phi)\right)\left[ e^{-1}\partial_{\mu}(eS_{a}{}^{\mu\nu})
    +T^{\rho}{}_{\mu a}S_{\rho}{}^{\mu\nu}
    -\frac{1}{4}e_{a}^{\,\,\,\nu}\mathcal{T}\right] \\
\notag    && - \frac{1}{2}e_{a}^{\,\,\,\nu}\left[\partial_{\mu}\phi\partial^{\mu}\phi-\frac{1}{\kappa}F_{\alpha \beta } F^{\alpha \beta }\right] +4 \frac{df(\phi)}{d\phi}(\partial_{\mu}\phi) S_{a}{}^{\mu\nu}\\
    && +\partial^{\nu}\phi\partial_{a}\phi
    -\frac{2}{\kappa}F^{\nu\lambda}F_{a\lambda}= 0\,,
\end{eqnarray} 
\begin{equation}\label{KGeq}
e^{-1}\partial_\nu(e g^{\mu\nu}\partial_\mu\phi) - f'(\phi) \mathcal{T} = 0  \,,
\end{equation}
\begin{equation}\label{Maxeq}
e^{-1}\partial_\nu\left(\frac{e}{\kappa}F^{\nu\mu}\right) = 0  \,.
\end{equation}

In order to obtain black hole solutions, we impose spherical symmetry and consider the following tetrad, which satisfies this symmetry in the Weitzenb\"{o}ck gauge \cite{Bahamonde:2022chq,Bahamonde:2021srr,Bahamonde:2022lvh} 
\begin{equation}
\resizebox{\columnwidth}{!}{$\displaystyle
e^{a}_{\ \mu} =
\begin{pmatrix}
0 & \frac{i}{\sqrt{B(r)}} & 0 & 0\\
i\sqrt{A(r)}\cos\varphi\sin\theta & 0 & -r\sin\varphi & -r\cos\theta\cos\varphi\sin\theta\\
i\sqrt{A(r)}\sin\theta\sin\varphi & 0 & r\cos\varphi  & -r\cos\theta \sin\theta\sin\varphi\\
i\sqrt{A(r)}\cos\theta & 0 & 0 & r\sin^2\theta
\end{pmatrix}\,.$}
\label{tetrad}
\end{equation}
This complex tetrad reproduces the metric in the standard static, spherically symmetric form in spherical coordinates
\begin{equation}\label{metric1}
ds^2 = A(r)dt^2 - \frac{1}{B(r)}dr^2 - r^2(d\theta^2 + \sin^2\theta d\varphi^2)\,.
\end{equation}
The functions $A,\, B,\, \phi,\, \Psi$ have only a radial dependence. For notational simplicity, this dependence will be omitted henceforth. A prime denotes differentiation with respect to $r$, while a dot denotes differentiation with respect to the scalar field $\phi$.
Using the tetrad given in Eq. (\ref{tetrad}), the torsion scalar defined in Eq. (\ref{scalar-T}) takes the form
\begin{equation}
\mathcal{T} = \frac{2 \left(r B A'+A (B+1)\right)}{r^2 A}\,.
    \label{torsionscalar}
\end{equation}
Substituting this expression into the field equations (\ref{eomst}) and contracting them with $e^{a}_{\,\,\, \sigma}$, we obtain the following independent components:
\begin{eqnarray}
\label{ett} 
E^t_{\;t} & \equiv & A \big[ 2 \left(r B'+B-1\right) (2 \kappa f(\phi) +1)+ 8 \kappa r B \phi ' \dot{f}(\phi) \nonumber \\
&&  + \kappa r^2 B \phi'^2 \big] + 2 r^2 B \Psi'^2 = 0 \,,\\[1em]
\label{err}  
E^r_{\;r} & \equiv &  A \big[ 2 (B-1) (2 \kappa f(\phi)+1) - \kappa r^2 B \phi '^2 \big]\nonumber \\
&& + 2 r B \left(A' (2 \kappa f(\phi)+1)+r \Psi'^2 \right) = 0\,,\\ [1em]
\notag 
E^\theta_{\;\theta} & \equiv & 4 \kappa A B \left(r A'+2 A\right) \phi' \dot{f}(\phi) - (2 \kappa f(\phi)+1) \big[ r B A'^2 \nonumber \\
&&  - A \big(r A' B' + 2 B \left(r A''+A'\right)\big) - 2 A^2 B' \big]  \nonumber \\
\label{eang} 
&& + 2 r A B \left(\kappa A \phi'^2-2\Psi'^2 \right) = 0 \,.
\end{eqnarray}
It should be noted that the angular components satisfy $E^\theta_{\;\theta} = E^\varphi_{\;\varphi}$.

Similarly, the modified Klein-Gordon equation (\ref{KGeq}) becomes
\begin{eqnarray}
E_{\phi} & \equiv & r B A' \left(4 \dot{f}(\phi)+r \phi '\right) + A \big[ r \big(\left(r B'+4 B\right) \phi'\nonumber \\
&& +2 r B \phi ''\big) + 4 (B+1) \dot{f}(\phi) \big] = 0 \label{ekg} \,.
\end{eqnarray}

Finally, the Maxwell equation (\ref{Maxeq}) can be integrated once, yielding
\begin{eqnarray}\label{emax}
   E^{\mu} & \equiv &   r^4 \Psi'^2 B -\kappa^2Q^2 A=0 \,,
\end{eqnarray}
where the integration constant $Q$ is interpreted as the electric charge of the black hole.

It is worth noting that when the scalar field is constant, $\phi=\phi_0$, and the coupling function satisfies $f(\phi_0)=f'(\phi_0)=0$, the field equations reduce to those of Einstein-Maxwell theory, admitting the RN solution. In particular, for the trivial scalar field configuration $\phi = 0$, the metric functions are given by
\begin{equation}
    A(r)=B(r)=1 -\frac{2M}{r}+\frac{Q^2}{r^2}\,,
\end{equation}
where $M$ and $Q$ denote the black hole mass and the electric charge, respectively. For $M^2 > Q^2$, this solution possesses two horizons: an outer event horizon and an inner Cauchy horizon, located at $r_{\pm}=M\pm \sqrt{M^2-Q^2}$.

\section{Decoupled Scalar Dynamics on fixed Reissner-Nordström Background}\label{decouplinglimit}

We investigate the dynamical formation of scalarized configurations and their stability within the decoupling limit. In this approximation, we solve the non-linear Klein-Gordon equation (\ref{KGeq}) while keeping the background geometry fixed to the RN solution.

\subsection{Time Evolution of the Scalar Field}

To investigate the onset of the instability triggered by the coupling functions $f_1(\phi)$ and $f_2(\phi)$, we numerically evolve the time-dependent Klein-Gordon equation (\ref{KGeq}), which can be written as
\begin{eqnarray}
  \notag  \partial^2_t \phi &=& h(r,q,r_+)\Bigl[n(r,q,r_+)\frac{df_i}{d\phi} + l(r,q,r_+)\partial_r \phi \\
    && + m(r,q,r_+)\partial^2_r \phi \Bigr] \,,
\end{eqnarray}
where the functions $h(r,q,r_+)$, $n(r,q,r_+)$, $l(r,q,r_+)$ and $m(r,q,r_+)$ are defined as
\begin{eqnarray}
         h(r,q,r_+) &=& \frac{(r-r_+) \left(r-q^2r_+\right)}{r^6}\,,\\ [1em]
          n(r,q,r_+) &=& 4 r^2 -2 q^2 r_+^2\,,\\[1em]
         l(r,q,r_+) &=& r^2(2r-r_+-q^2r_+)\,,\\ [1em]
         m(r,q,r_+) &=& r^2(r-r_+)(r-q^2r_+)\,.
\end{eqnarray}
Here, $q\equiv Q/r_+$ denotes the reduced electric charge. To solve this equation numerically in the time domain, we employ the method of lines. Spatial derivatives are discretized using finite differences, and the resulting system of ordinary differential equations is evolved over time using a fourth-order Runge-Kutta scheme \cite{Baumgarte:2021skc}. Regarding the boundary conditions, we impose the Sommerfeld radiation condition, which ensures that the scalar field perturbations propagate as ingoing waves at the black hole horizon and as outgoing waves at spatial infinity. In our numerical implementation, the computational domain in the radial direction extends over the interval $[x_{\text{in}}, x_{\text{out}}]$. To avoid the coordinate singularity at the event horizon, the inner boundary is placed slightly outside it, at $x_{\text{in}} = 1.5 r_+$. The outer boundary, in turn, is located sufficiently far from the black hole, at $x_{\text{out}} = 300.0$, to properly capture the asymptotic behavior of the scalar field and  minimize spurious reflections by imposing the outgoing wave condition. The entire spatial domain is discretized using a uniform grid with $N = 300$ points.

As mentioned above, the RN solution is stable against linear scalar perturbations. Therefore, a finite amplitude perturbation is needed to activate the nonlinear terms in the scalar field equation and potentially trigger the dynamical formation of a scalarized black hole configuration. To this end, we consider a Gaussian wave packet with finite amplitude $A$ as the initial scalar field perturbation
\begin{eqnarray}
        \phi(t=0,r) &=& A \exp\left(-\frac{(r-\overline{r})^2}{2\sigma^2}\right)\,,\\ 
        \partial_t\phi(t=0,r)&=&0\,.
\end{eqnarray}
The wave packet is initially localized outside the event horizon, with its peak located at $\bar{r} = 10 r_+$ and a width parameter $\sigma = 0.5$. We set the event horizon radius to $r_+ = 1$, and evolve the system up to a maximum time of $t_{\text{max}} = 250$. To ensure numerical stability during the time evolution, we impose the Courant-Friedrichs-Lewy (CFL) condition with a Courant factor of $0.5$, which sets the time step to $\Delta t = 0.5 \Delta x$. Finally, the time evolution of the scalar field is sampled every $t_{\text{check}} = 0.2$ and extracted at a fixed observation radius of $r_{\text{obs}} = 20 r_+$.

\begin{figure}[H]
\begin{center}
\includegraphics[width=0.40\textwidth]{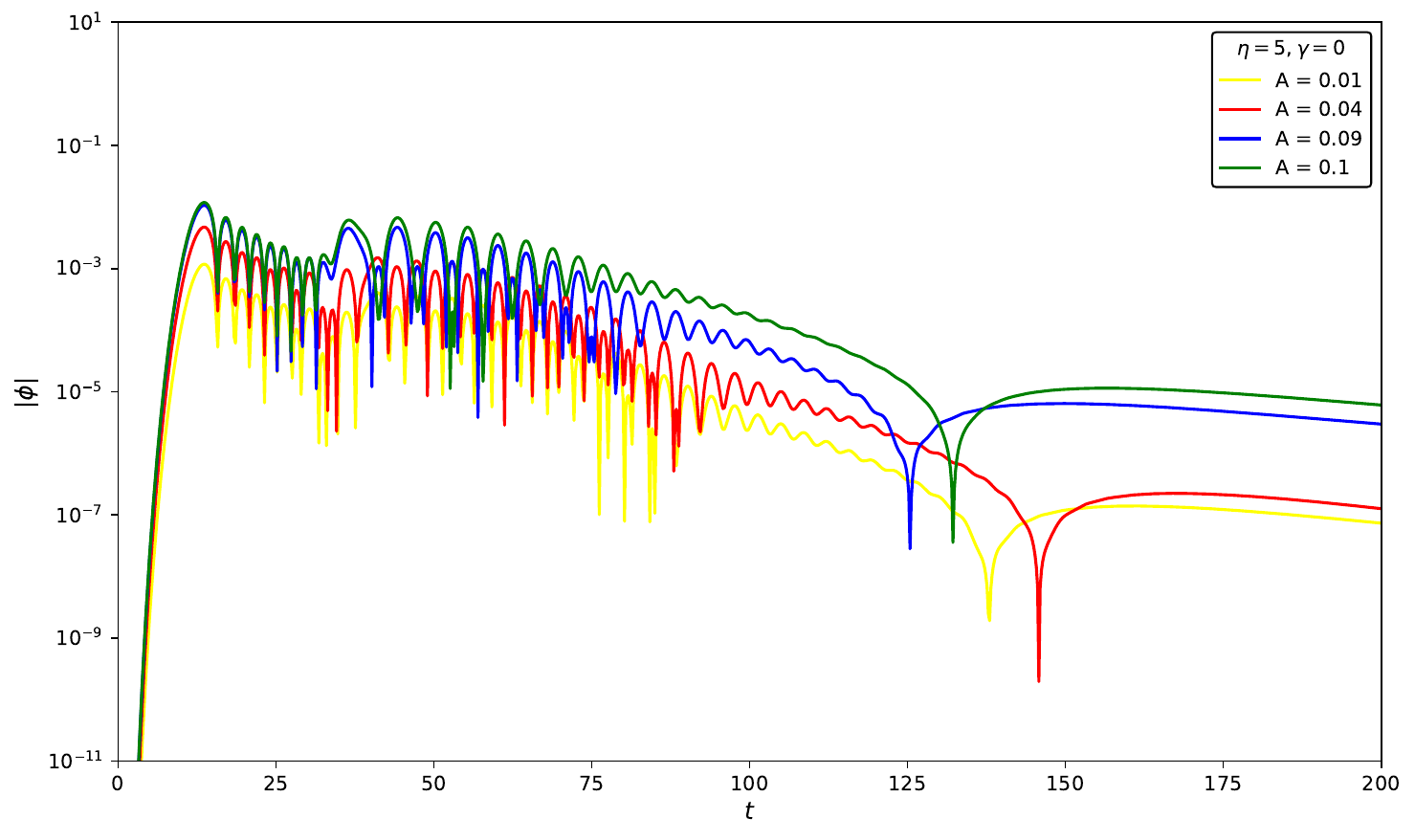}
\includegraphics[width=0.40\textwidth]{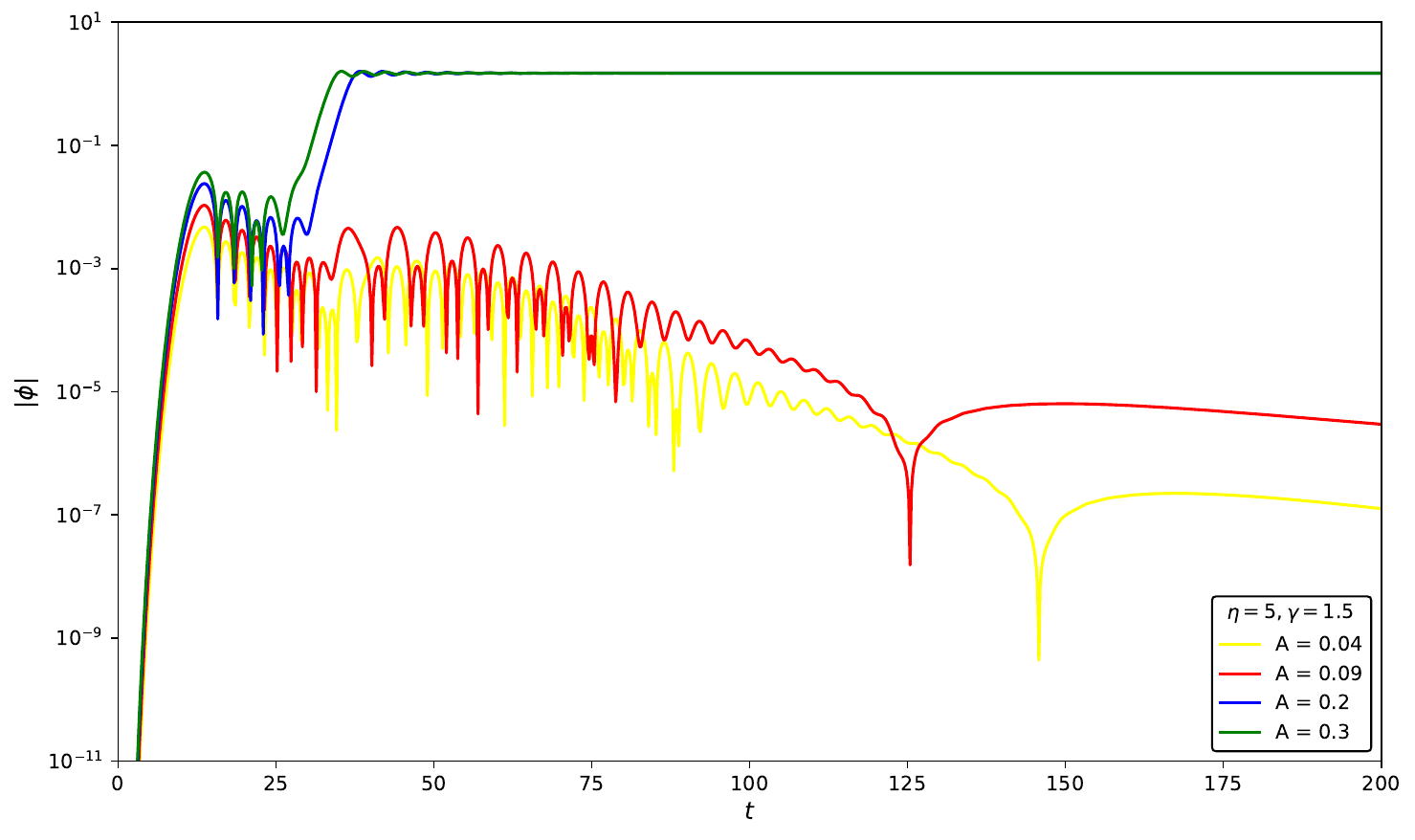}
\end{center}
\caption{Time evolution of the scalar field on the RNBH background for $q=0.5$ and $r_+=1$ (i.e., $M=0.625$). The top and bottom panels display the results for the coupling functions $f_1(\phi)$ and $f_2(\phi)$, respectively, using $\eta=5$ and $\gamma=1.5$.}
\label{evokgeq}
\end{figure}

By evolving the nonlinear scalar field equation in time on the RN background, we observe distinct dynamical behaviors depending on the choice of the coupling function. For $f_2(\phi) = \eta \phi^4 - \gamma\phi^6$ (with $\gamma > 0$), we identify a threshold in the initial perturbation amplitude. For amplitudes below this threshold, the scalar field undergoes standard exponential decay. Above the threshold, the scalar field initially grows but eventually saturates and approaches a time-independent, nontrivial configuration, providing evidence for dynamical stabilization, as shown in Fig. \ref{evokgeq}. In contrast, for the purely quartic coupling $f_1(\phi) = \eta\phi^4$ (i.e., $\gamma=0$), no stable scalarized phases are dynamically formed. While the scalar field undergoes standard decay for small initial amplitudes, a large initial perturbation causes the field to diverge within a short timescale. This unbounded growth signifies a runaway instability, preventing the dynamic formation of stable scalarized black holes. Nevertheless, it is worth mentioning that although the coupling function $f_1(\phi)$ does not lead to dynamically stable phases, the full system of equations (\ref{ett})--(\ref{emax}) does yield stationary solutions.

\subsection{Effective Potential}

To gain further insight into the scalar-field dynamics and the role played by the coupling functions $f_i(\phi)$, we rewrite the time-independent Klein-Gordon equation in terms of an effective scalar potential. For the RN background, this potential is given by
\begin{equation}
    V_{\text{eff}}^{(i)}(r, \phi) = \frac{2(q^2r_+^2 - 2r^2)}{r^4} f_i(\phi) \,.
\end{equation}
The evolutionary states and the stability of the scalar field can be physically interpreted through the shape of this potential. The local extrema are obtained by solving $d V_{\text{eff}}^{(i)}/d\phi = 0$. For the purely quartic coupling $f_1(\phi) = \eta \phi^4$, the only extremum is located at the trivial configuration:
\begin{eqnarray}
   \phi_c = 0 \,. 
\end{eqnarray}
Meanwhile, for the coupling function $f_2(\phi) = \eta \phi^4 - \gamma \phi^6$, the potential develops additional non-trivial local extrema given by
\begin{equation}
    \phi_c = \left\{0, \pm \sqrt{\frac{2\eta}{3\gamma}} \right\} \,.
\end{equation}

\begin{figure}[H]
\begin{center}
\includegraphics[width=0.40\textwidth]{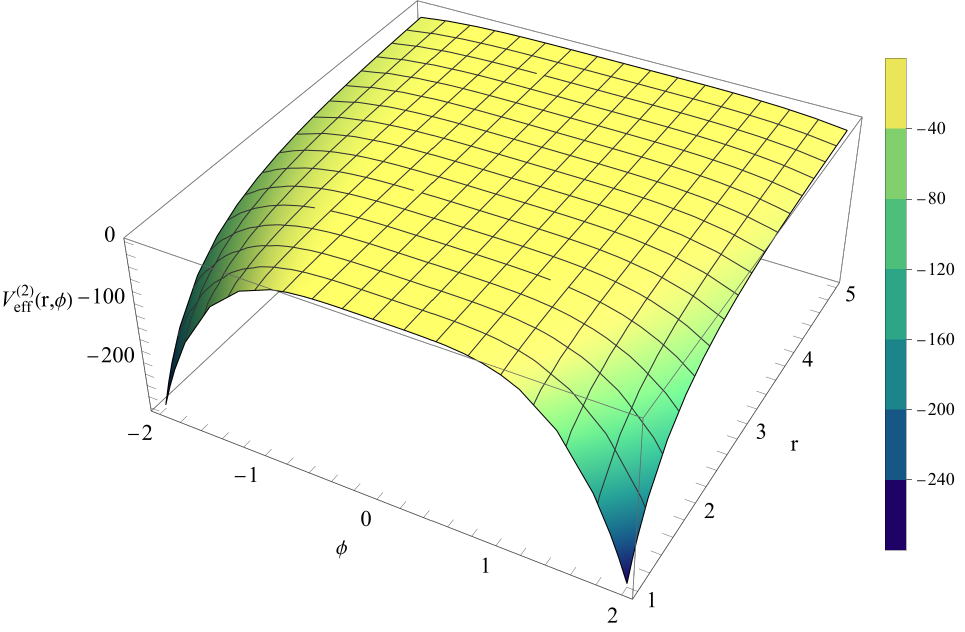}
\includegraphics[width=0.40\textwidth]{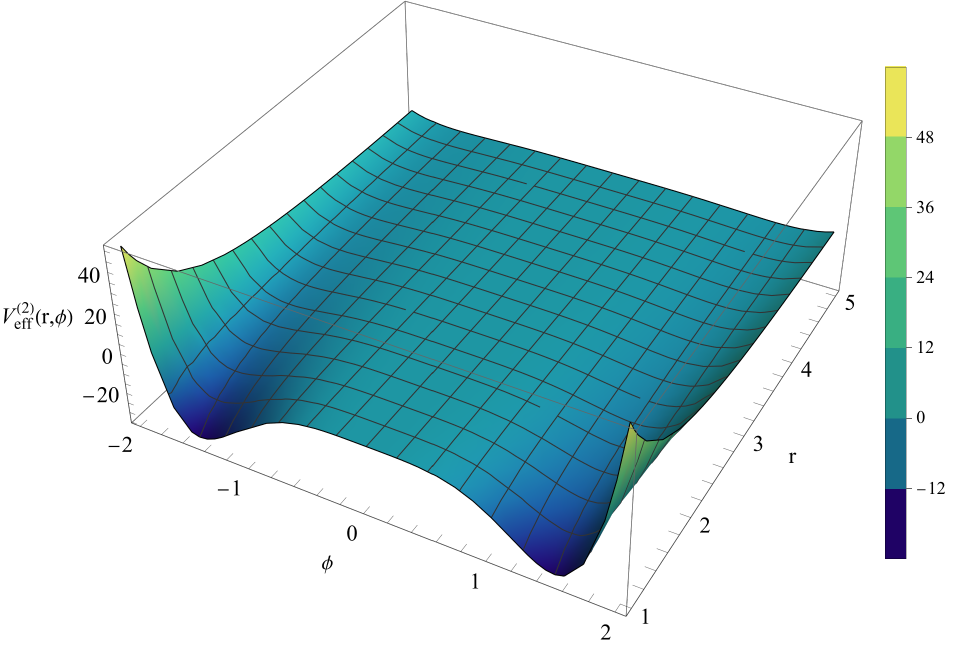}
\end{center}
\caption{Effective potential for a scalar field on the RN black hole background with $q=0.5$ and $r_+=1$ (corresponding to $M=0.625$). The top and bottom panels show the effective potentials associated with the coupling functions $f_1(\phi)$ and $f_2(\phi)$, respectively, for $\eta=5$ and $\gamma=1.5$.}
\label{effpotential}
\end{figure}

For the quartic coupling $f_1(\phi) = \eta\phi^4$, the trivial configuration $\phi_c = 0$ corresponds to a local maximum of the effective potential, as shown in the top panel of Fig. \ref{effpotential}. Since the potential contains no higher-order terms that can bound it from below, it decreases without bound as $\phi$ increases. Consequently, the scalar field undergoes a nonlinear runaway, cascading down this unbounded slope. This explains the divergence observed in our time evolution results, indicating that the purely quartic coupling does not
provide a mechanism for nonlinear saturation of the
scalar field in the decoupling limit.
In contrast, the inclusion of the sixth-order term $\gamma\phi^6$ in the coupling function $f_2(\phi)$ alters the landscape of the effective potential, as illustrated in the bottom panel of Fig. \ref{effpotential}. This term provides a repulsive feedback that balances the attractive $\eta\phi^4$ interaction at larger values of $\phi$. As a result, the potential develops a double-well structure characterized by stable potential wells located at the non-trivial extrema $\phi_c = \pm \sqrt{2\eta / 3\gamma}$. Since the effective potential depends explicitly on the radial coordinate, these wells are most pronounced near the black hole horizon and become progressively shallower towards spatial infinity. When a sufficiently large initial perturbation surmounts the central barrier at $\phi=0$, the scalar field becomes permanently trapped within one of these local minima. The system subsequently settles into a time-independent configuration, corresponding to the saturation plateau observed during the dynamic evolution. This mechanism provides a physical interpretation
of the nonlinear saturation observed in the
time-domain evolution and the emergence of
nontrivial stationary scalar configurations
in the decoupling limit.

The effective potential, therefore, provides a physical interpretation of the dynamical behavior observed in Fig. \ref{evokgeq}. For the purely quartic coupling, the potential is unbounded from below and does not provide a mechanism for nonlinear saturation. In contrast, the sixth-order coupling term introduces two nontrivial local minima, which can limit the growth of finite-amplitude scalar perturbations. This behavior is consistent with the numerical evolution, in which sufficiently large perturbations approach a nontrivial stationary scalar configuration. Taken together, the time evolution and the effective potential illustrate how higher-order scalar-torsion couplings can dynamically stabilize the scalar field in the decoupling limit.

\section{Scalarization of the Reissner-Nordström Black Hole}\label{numeric}

In this section, we construct nonlinear scalarized black hole solutions by numerically solving the full system of field equations (\ref{ett})-(\ref{emax}).

To facilitate numerical integration and ensure asymptotic flatness, we introduce a compactified radial coordinate $z=1-r_+/r$ and redefine the metric functions, the scalar field, and the electric potential as follows:
\begin{eqnarray}\label{IRA}
    A(z) = z a(z)\,, \quad B(z)=zb(z)\,,\\ 
    \phi(z)=(1-z)\tilde{\phi}(z)\,, \quad \Psi(z)=z\psi(z)\,.
\end{eqnarray}

The differential equations are solved using the shooting method \cite{stoer1980introduction, press2007numerical}. Specifically, the equations are integrated outward from the horizon ($z = 0$) and inward from spatial infinity ($z = 1$), with the two solutions required to match at an intermediate point $z_{\text{match}} = 0.5$. The numerical integration is performed with an internal precision parameter $\epsilon = 10^{-8}$, while the shooting parameter procedure employs a tolerance of $\delta = 10^{-7}$. These settings ensure that the mismatch between the two integrated solutions at $z_{\text{match}}$ is smaller than $10^{-5}$.

To construct the near horizon boundary conditions, we expand the metric functions, the scalar field, and the electric potential in Taylor series about the event horizon ($z=0$). Substituting these expansions into the field equations (\ref{ett})-(\ref{emax}), we obtain
\begin{eqnarray}\label{IR}
\notag    a(z) &=& a_H + \sum_{n=1}^{10}a_{Hn}(\phi_H,a_H,\gamma,\eta,q)z^n\,,\\
\notag   b(z) &=& b_H + \sum_{n=1}^{10}b_{Hn}(\phi_H,a_H,\gamma,\eta,q)z^n\,, \\
\notag    \tilde{\phi}(z) &=& \phi_H +\sum_{n=1}^{10}\phi_{Hn}(\phi_H,a_H,\gamma,\eta,q)z^n\,, \\
   \notag  \psi(z) &=& \kappa\sqrt{\frac{a_H}{b_H}}q+\sum_{n=1}^{10}\psi_{Hn}(\phi_H,a_H,\gamma,\eta,q)z^n\,,
\end{eqnarray}
where $b_H=1-\frac{q^2\kappa^2}{1+2\kappa\eta\phi_H^4-2\kappa\gamma\phi_H^6}$. Similarly, by expanding the field equations around spatial infinity ($z \rightarrow 1$), we obtain
\begin{eqnarray}\label{IRB}
\notag    a(z) &=& 1 + \sum_{n=1}^{10}a_{n}(\phi_0,b_1,\psi_0,\gamma,\eta,q)(1-z)^n\,,\\ 
\notag     b(z) &=& 1 + b_1(1-z)+\sum_{n=2}^{10}b_{n}(\phi_0,b_1,\psi_0,\gamma,\eta,q)(1-z)^n\,, \\
 \notag   \tilde{\phi}(z) &=& \phi_0 +\sum_{n=1}^{10}\phi_{n}(\phi_0,b_1,\psi_0,\gamma,\eta,q)(1-z)^n\,,\\
 \notag   \psi(z) &=& \psi_0+\sum_{n=1}^{10}\psi_{n}(\phi_0,b_1,\psi_0,\gamma,\eta,q)(1-z)^n\,.
\end{eqnarray}

Asymptotic flatness is imposed by requiring $a_{0}= 1$ and $b_{0} = 1$. The coefficient $b_1$ is related to the ADM mass $\mathcal{M}$ of the scalarized black hole through $\mathcal{M}=(1-b_1)r_+/2$. The parameter $\psi_0$ represents the electric potential at spatial infinity, which, for the RN solution, corresponds to the reduced charge, i.e., $\psi_{0}^{RN}=q\equiv Q/r_+$. The asymptotic scalar field behaves as $\phi(r) \sim \phi_0 r_+/r$, so that $\phi_0 r_+$ corresponds to the scalar charge. Finally, we consider solutions with a non-vanishing scalar field at the event horizon (i.e., $\phi_H \neq 0$), thereby ensuring a non-trivial scalar field profile. The regularity of the solutions is imposed through the near horizon boundary conditions.

The solutions of the full system are characterized by eight parameters: $\{\phi_H, a_H, \phi_0, b_1, \psi_0, q, \eta, \gamma\}$. To explore the parameter space, we fix the reduced electric charge to $q=0.5$. For given values of $\eta$ and $\gamma=0$, bounded solutions satisfying the asymptotic condition $\phi(z \to 1) \to 0$ are obtained only for specific discrete combinations of the parameters $\{\phi_H, a_H, \phi_0, b_1, \psi_0\}$. In this case, we find a finite set of solutions, which can be represented as $\{(\phi_H, a_H, \phi_0, b_1, \psi_0; n)\}_{n=0}^{n_{\text{max}}}$. Here, $n$ denotes the number of nodes in the scalar field profile, while $n_{\text{max}}$ is the maximum number of nodes allowed for a given value of $\eta$. Notably, the solutions with the maximum number of nodes are degenerate, in the sense that distinct parameter configurations give rise to scalar fields profiles with the same number of nodes.


For fixed $\eta$, switching on the coupling $\gamma > 0$ gives rise to additional branches of scalarized solutions, significantly modifying the discrete set of configurations. This structural evolution is summarized in Table \ref{nodes_comparison}, which reports the number of solutions (NoS) and their corresponding node numbers for two representative values, $\eta=5$ and $\eta=30$. To distinguish the origin of these branches, the NoS column uses an additive notation (e.g., $3+2$), where the first integer denotes the number of solutions already present in the purely quartic case ($\gamma=0$), while the second counts additional solutions found for $\gamma>0$. As observed in the table, for small values of $\gamma$, additional degenerate solutions are introduced. As $\gamma$ increases, this degeneracy is eventually broken, thereby increasing the maximum number of nodes $n_{\text{max}}$. However, in the strong coupling regime where $\gamma \gg \eta$, the higher-node branches are systematically suppressed until only the fundamental solution ($n=0$) remains. Ultimately, for sufficiently large $\gamma$, even this fundamental branch disappears, resulting in a complete absence of scalarized solutions. 
\begin{table}[H]
\centering 
\scriptsize
\begin{tabular}{|c|c|c||c|c|c|}
\hline
\multicolumn{3}{|c||}{$\boldsymbol{\eta = 5}$} & \multicolumn{3}{c|}{$\boldsymbol{\eta = 30}$} \\
\hline
$\boldsymbol{\gamma}$ & \textbf{NoS} & \textbf{Nodes} & $\boldsymbol{\gamma}$ & \textbf{NoS} & \textbf{Nodes} \\
\hline
$0$    & 3   & 0, 1, 1       & $0$    & 4 & 0, 1, 2, 2 \\
\hline
$0.25$ & 3+2 & 0, 1, 1, 1, 2 & $0.5$  & 4+3 & 0, 1, 2, 2, 1, 1, 2 \\
\hline
$0.5$  & 2+2 & 0, 1, 2, 3 & $1.5$  & 4+3 & 0, 1, 2, 2, 2, 3, 4 \\
\hline
$1.5$  & 2+2 & 0, 1, 2, 3    & $9$    & 4+5 & 0, 1, 2, 2, 2, 3, 4, 5, 6 \\
\hline
$5$    & 1+2 & 0, 1, 2       & $15$   & 3+4 & 0, 1, 2, 3, 4, 5, 6 \\
\hline
$10$   & 1+1 & 0, 1          & $30$   & 3+3 & 0, 1, 2, 3, 4, 5 \\
\hline
$23$   & 1   & 0             & $830$  & 1 & 0 \\
\hline
$89$   & 0   & --            & $3300$ & 0 & -- \\
\hline
\end{tabular}
\caption{Number of solutions (NoS) and corresponding node numbers ($n$) for different values of $\gamma$, evaluated at $\eta=5$ and $\eta=30$.}
\label{nodes_comparison}
\end{table}

\begin{figure}[H]
  \begin{center}
    \includegraphics[width=0.4\textwidth]{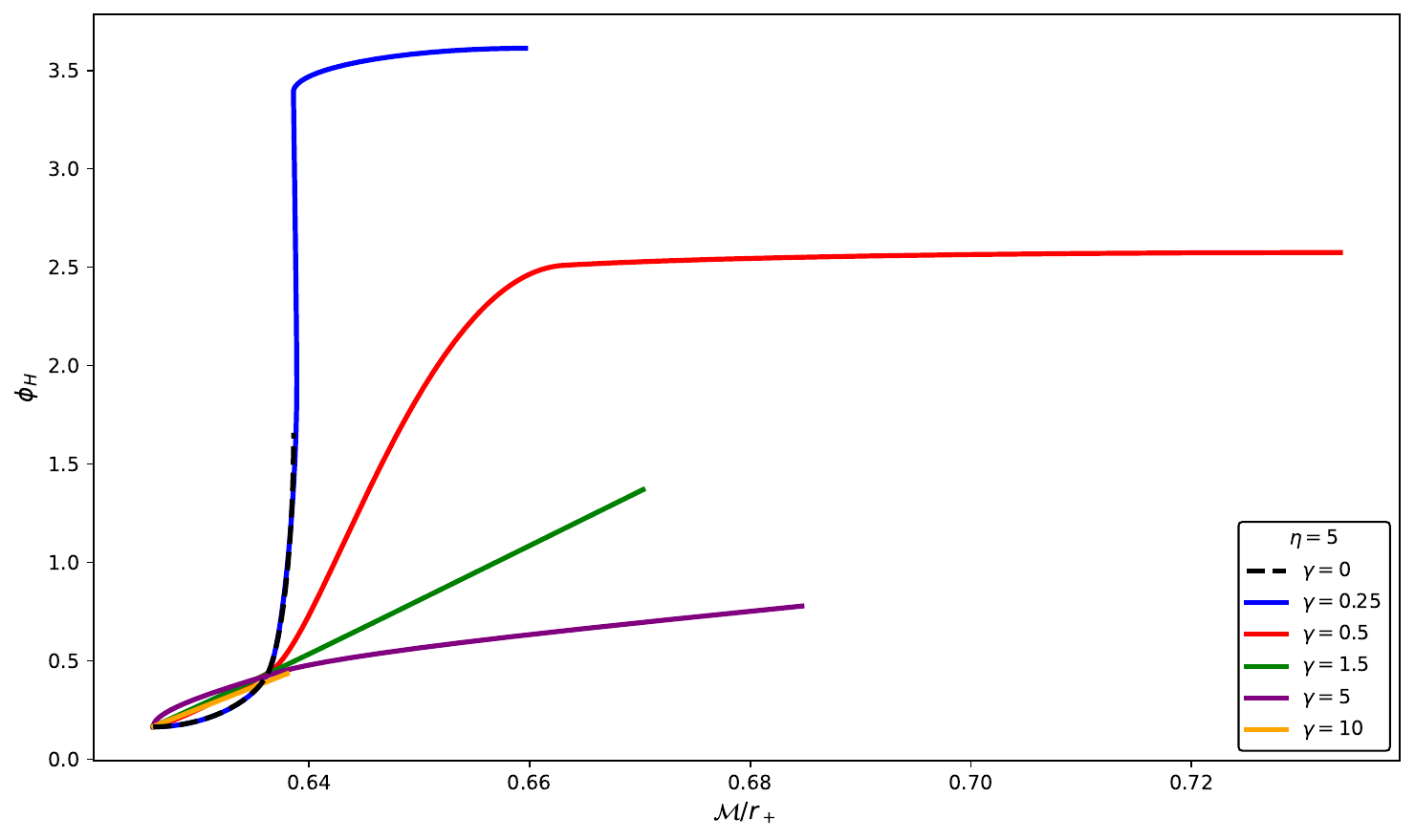}
    \includegraphics[width=0.4\textwidth]{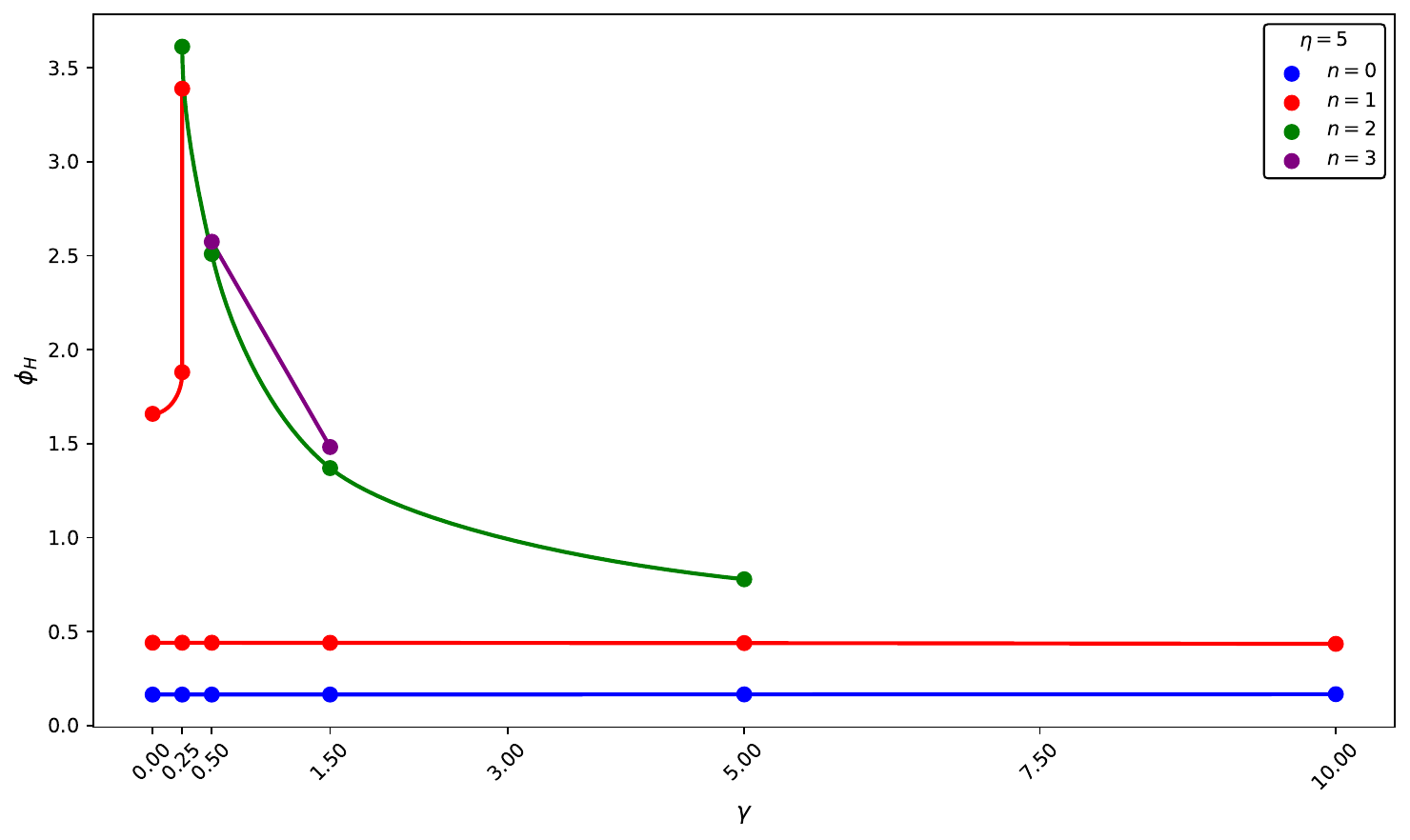}
  \end{center}
  \caption{Bifurcation diagrams for scalarized black hole solutions with $\eta=5$ and $q=0.5$. Top panel: Domain of existence showing $\phi_H$ as a function of $\mathcal{M}/r_+$. Bottom panel: $\phi_H$ as a function of $\gamma$, visualizing the breaking of degeneracy at weak couplings and the suppression of higher-node branches ($n \ge 1$) at strong couplings.}
  \label{bifurcation_diagrams}
\end{figure}

The macroscopic implications of this degeneracy and its evolution are best visualized through the bifurcation diagrams in Fig. \ref{bifurcation_diagrams}. The bottom panel explicitly demonstrates the breakup of degeneracy: multiple branches with the same node number are separated as $\gamma$ increases, whereas higher-node solutions are eventually suppressed. The top panel illustrates the domain of existence, showing how the emergence of the scalar field at the horizon ($\phi_H$) drives a deviation in the reduced ADM mass ($\mathcal{M}/r_+$) from the RN background.

To gain a deeper understanding of the physical properties of these scalarized branches, it is necessary to examine their behavior at the event horizon and at spatial infinity. Because the qualitative physical behavior in both regimes is completely analogous, our analysis will focus exclusively on the case of $\eta=5$. The detailed physical parameters for these solutions for various values of $\gamma$, as well as the corresponding results for case $\eta=30$, are tabulated in Tables \ref{parametersolutions} and \ref{parametersolutions_eta30} of Appendix \ref{Solutions}.

Having established the parameters, we now examine the radial profiles of these configurations. The following figures illustrate the behavior of the scalar field $\phi(z)$, the metric deviations $\delta A(z)$ and $\delta B(z)$, and the dimensionless torsion scalar deviation $r_+^2\delta\mathcal{T}(z)$ for $\eta=5$.

\begin{figure}[H]
\begin{center}
\includegraphics[width=0.4\textwidth]{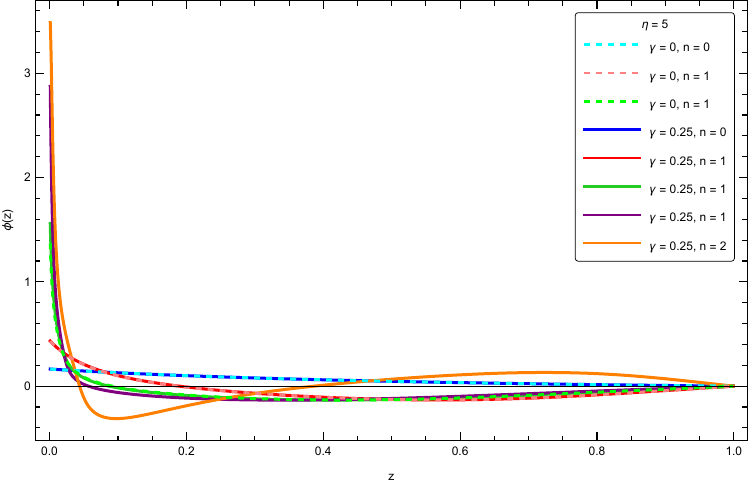}
\includegraphics[width=0.4\textwidth]{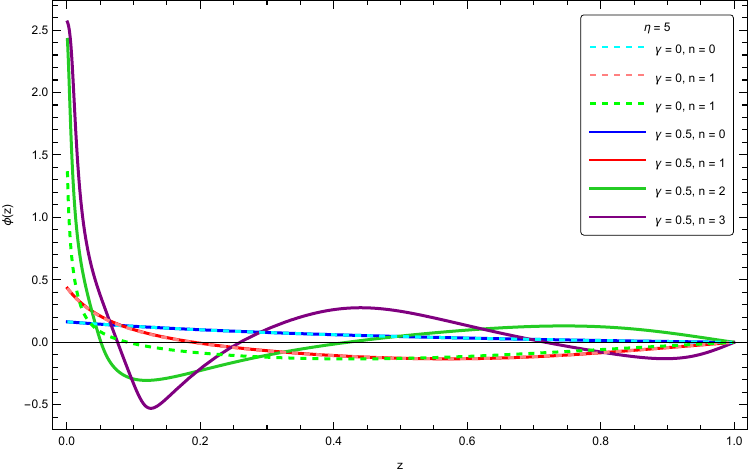}
\end{center}
\caption{Radial profiles of the scalar field $\phi(z)$
for scalarized black hole solutions with $\eta=5$ and
$q=0.5$. The top panel compares the solutions for
$\gamma=0$ and $\gamma=0.25$, while the bottom panel
shows the corresponding comparison for $\gamma=0$
and $\gamma=0.5$. The different curves illustrate
the emergence of additional scalarized branches
and the changes in their nodal structure induced
by the sixth-order coupling term.
}
\label{plotsphi1}
\end{figure}

\begin{figure}[H]
\begin{center}
\includegraphics[width=0.4\textwidth]{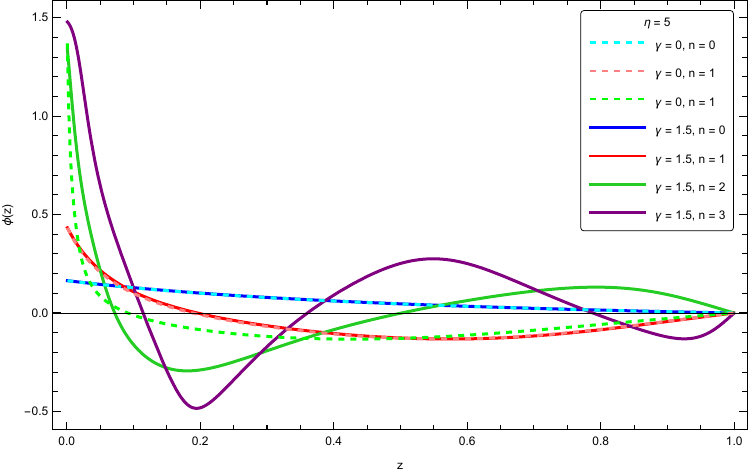}
\includegraphics[width=0.4\textwidth]{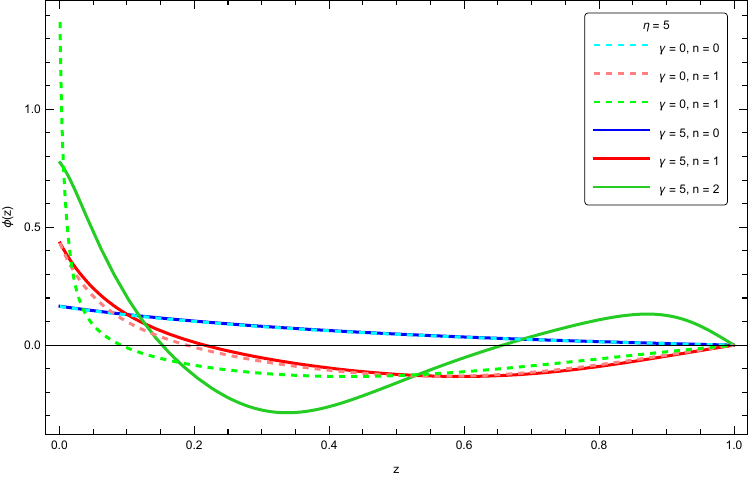}
\end{center}
\caption{
Radial profiles of the scalar field $\phi(z)$
for scalarized black hole solutions with $\eta=5$ and
$q=0.5$. The top panel compares the solutions for
$\gamma=0$ and $\gamma=1.5$, while the bottom panel
shows the corresponding comparison for $\gamma=0$
and $\gamma=5$. The profiles illustrate the evolution
of the nodal branches as the sixth-order coupling
increases, including the progressive suppression
of higher-node solutions.
}
\label{plotsphi2}
\end{figure}

Figs. \ref{plotsphi1} and \ref{plotsphi2} display the radial evolution of the scalar field. To highlight the impact of the coupling, each panel contrasts the reference solutions at $\gamma=0$ with those for specific non-zero values ($\gamma \in \{0.25, 0.5, 1.5, 5\}$). This comparison explicitly visualizes the evolution of the nodal structure and how the degeneracy of the $\gamma=0$ solutions is broken as $\gamma$ increases. Specifically, it can be seen that while the fundamental nodeless solution remains nearly identical across different couplings, the initially degenerate one-node solutions (represented by the red and dashed-pink curves) progressively differentiate from each other as $\gamma$ grows. 

Furthermore, as physically required, all numerical solutions are regular and strictly satisfy the asymptotic boundary condition, decaying to the vacuum $\phi \to 0$ as $z \to 1$.

\begin{figure}[H]
\begin{center}
\includegraphics[width=0.4\textwidth]{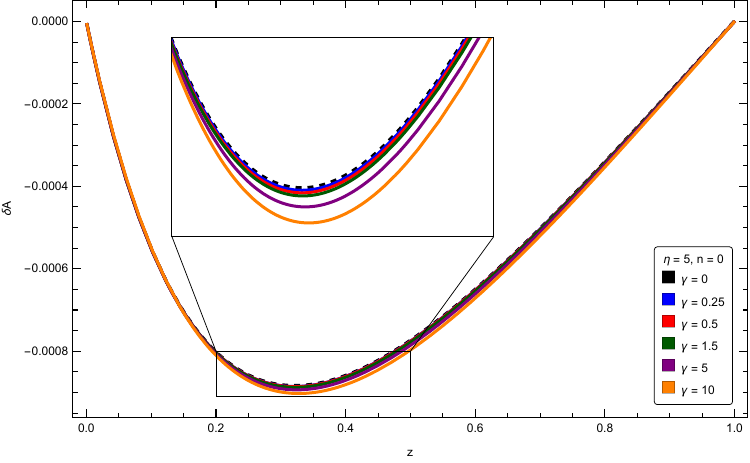}
\includegraphics[width=0.4\textwidth]{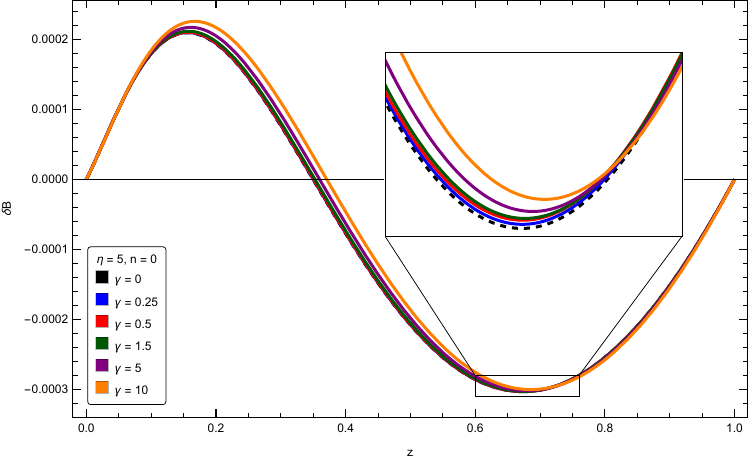}
\end{center}
\caption{The behavior of the metric deviations $\delta A = A_{sc}-A_{RN}$ (top panel) and $\delta B = B_{sc}-B_{RN}$ (bottom panel) as a function of $z$ for the fundamental $n=0$ solutions. The different curves correspond to various values of $\gamma$ evaluated at a fixed $\eta=5$.}
\label{delta_A_B}
\end{figure}

In Fig. \ref{delta_A_B}, we analyze the spatial behavior of the metric deviations $\delta A$ and $\delta B$ for the fundamental nodeless solutions ($n=0$) relative to the RN background. As observed in the top panel, the deviation $\delta A$ remains strictly negative across the entire domain, from the event horizon ($z=0$) to spatial infinity ($z=1$). This negative deviation indicates that the backreaction of the scalar field uniformly modifies the temporal metric component, effectively deepening the gravitational potential well outside the black hole without disrupting the asymptotic flatness.

On the other hand, the bottom panel reveals a spatially dependent transition for the radial metric deviation $\delta B$. In the near-horizon regime (from $z=0$ up to $z \approx 0.35$), $\delta B$ is distinctly positive, reflecting the strong localized deformation of the geometry caused by the concentration of the scalar field's energy density. However, as one moves further outward ($z \gtrsim 0.35$), $\delta B$ changes sign and becomes negative, eventually decaying to zero. This transition explicitly demonstrates that scalarization is fundamentally a near-horizon phenomenon; the scalar hair is strongly anchored in the vicinity of the black hole, and its geometrical backreaction progressively weakens as it approaches the asymptotic region, where the standard RN vacuum is smoothly recovered.

\begin{figure}[H]
\begin{center}
\includegraphics[width=0.4\textwidth]{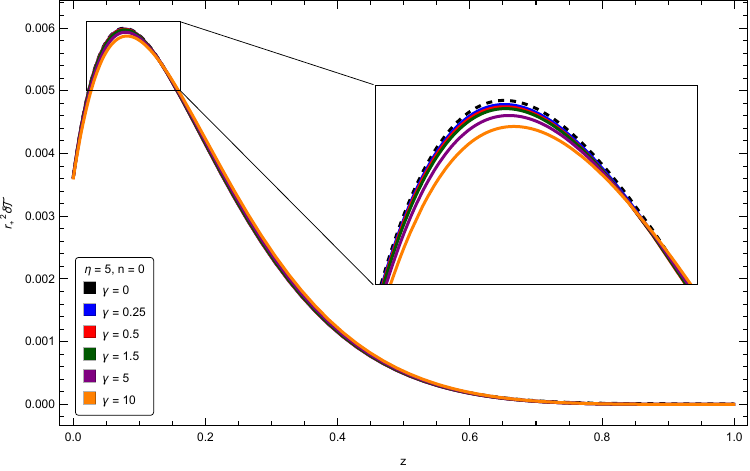}
\end{center}
\caption{The behavior of the torsion scalar deviation $r_+^2\delta \mathcal{T} = r_+^2 \mathcal{T}_{sc}-r_+^2\mathcal{T}_{RN}$ as a function of $z$ for the fundamental $n=0$ solutions. The different curves correspond to various values of $\gamma$ evaluated at a fixed $\eta=5$.}
\label{delta_T}
\end{figure}

Finally, we analyze the scalarization in the teleparallel spacetime structure, depicted in Fig. \ref{delta_T}. This figure illustrates the dimensionless torsion scalar deviation, $r_+^2\delta \mathcal{T} = r_+^2\mathcal{T}_{sc} - r_+^2\mathcal{T}_{RN}$, as a function of the compactified radial coordinate $z$ for the fundamental nodeless solutions ($n=0$). We observe that this deviation remains strictly positive throughout the entire exterior spacetime region. Interestingly, the maximum geometrical deviation does not occur exactly at the event horizon ($z=0$), but instead is displaced slightly outward, peaking near $z \approx 0.1$. This indicates that the most intense modification of the spacetime torsion by the scalar hair takes place in the immediate vicinity outside the black hole rather than on the horizon itself. Beyond this peak, the deviation decays to zero as $z \to 1$. It is worth highlighting that this asymptotic decay is fully consistent with the required boundary conditions; if one computes the absolute torsion scalar of the scalarized solution ($\mathcal{T}_{sc}$) rather than the deviation, it correctly recovers the standard asymptotically flat behavior, where the torsion strictly vanishes at spatial infinity.

\section{Final Remarks}
\label{conclusion}

In this work, we have investigated the scalar-torsion-Maxwell gravity within the framework of teleparallel gravity, where a scalar field is non-minimally coupled to the Weitzenböck invariant. By employing coupling functions specifically designed to prevent tachyonic instabilities, we utilized the numerical shooting method to demonstrate the formation of black hole solutions endowed with scalar hair.

Our analysis of the nonlinear scalar dynamics in the
decoupling limit reveals distinct behaviors for the two
polynomial coupling functions. For the coupling function
$f_1(\phi)=\eta\phi^4$, the effective potential is
unbounded from below and does not provide a mechanism
for nonlinear saturation, allowing sufficiently large
scalar perturbations to undergo runaway growth.
Although stationary scalarized solutions exist for
$\gamma=0$ in the fully coupled system, their dynamical
stability remains to be investigated. In contrast,
the coupling function $f_2(\phi)=\eta\phi^4-\gamma\phi^6$
allows finite-amplitude scalar perturbations to evolve
toward nontrivial stationary configurations in the
decoupling limit. This dynamical stabilization is
consistent with the effective potential, where the
sixth-order coupling term generates a double-well
structure with two nontrivial local minima that provide
a mechanism for nonlinear saturation.

It is worth noting that qualitatively similar behavior has been reported in curvature-based Einstein-scalar-Gauss-Bonnet gravity, where polynomial coupling functions have been shown to play a crucial role in nonlinear scalarization \cite{Zou:2024wsk}. In particular, a purely quartic coupling admits stationary scalarized solutions but does not lead to dynamically stabilized configurations, whereas the inclusion of negative sixth- or eighth-order terms allows finite-amplitude scalar perturbations to evolve toward nontrivial stationary states. This behavior is understood through the effective potential, whose higher-order terms provide a mechanism for nonlinear saturation. Our results reveal an analogous mechanism in scalar-torsion-Maxwell teleparallel gravity, where the scalar field couples to the torsion invariant rather than to the Gauss-Bonnet curvature invariant. This suggests that the nonlinear saturation mechanism is not restricted to a particular geometric formulation of gravity but depends crucially on the structure of the scalar coupling function.

Our results further demonstrate that the inclusion of the sixth-order coupling term significantly modifies the structure of the scalarized black hole solutions. Beyond its role in the nonlinear saturation of scalar perturbations, this term alters the multiplicity of the nodal branches, allowing additional scalarized configurations to emerge for intermediate values of the coupling parameter. However, as $\gamma$ increases further, higher-node branches are progressively suppressed, and for sufficiently large values of $\gamma$, even the fundamental nodeless branch ceases to exist. These results highlight the important role of higher-order scalar-torsion couplings in both the nonlinear scalar dynamics and the existence and structure of scalarized black hole solutions.

Furthermore, we have analyzed the backreaction of the scalar field on the spacetime geometry by comparing the metric functions and the torsion scalar of the scalarized solutions with those of the RN background. For the fundamental nodeless branches, we find that the scalar field induces deviations in both metric functions, while the torsion scalar exhibits a positive deviation whose maximum occurs slightly outside the event horizon. These deviations gradually vanish toward spatial infinity, consistent with the asymptotically flat character of the solutions. This behavior provides a geometric characterization of nonlinear scalarization within the teleparallel framework, complementing the dynamical analysis performed in the decoupling limit.



Looking forward, it would be interesting to investigate the linear stability of the fully backreacted scalarized solutions and explore their thermodynamic properties to determine whether these configurations are entropically and thermodynamically favored over the RN background. Furthermore, studying the dependence of the scalarized branches on the electric charge would provide a deeper understanding of the role of the Maxwell field in the nonlinear scalarization mechanism. Alongside this, our aim is to investigate the optical properties of these solutions by calculating the angular shadows of the scalarized black holes \cite{Wang:2024lte}. Comparing these predictions with the Event Horizon Telescope observations of black holes M87* \cite{EventHorizonTelescope:2019ggy} and Sagittarius A* \cite{EventHorizonTelescope:2022xqj} could provide observational constraints on the coupling parameters $(\eta,\gamma)$ and help assess the phenomenological viability of these scalarized configurations.


\begin{acknowledgments}

Y. V. acknowledges the financial support of DIDULS/ULS, through the project No PR25538511.

\end{acknowledgments}

\newpage

\appendix
\section{Parameters of Scalarized Black Hole Solutions}
\label{Solutions}

In the following tables, we show the values of the parameters $\gamma$, $n$, $\phi_H$, $a_H$, $\phi_0$, $b_1$, and $\psi_0$ for both the Reissner-Nordstr\"{o}m and scalarized black hole solutions, for $\eta=5$ in Table \ref{parametersolutions} and $\eta=30$ in Table \ref{parametersolutions_eta30}.

\begin{widetext}

\begin{table*}[h!]
  \centering
  \scriptsize
  \begin{tabular}{|c|c|c|c|c|c|c|c|}
    \hline
    \textbf{Type of BH} & $\boldsymbol{\gamma}$ & \textbf{Nodes} & $\boldsymbol{\phi_H}$ & $\boldsymbol{a_H}$ & $\boldsymbol{\phi_0}$ & $\boldsymbol{b_1}$ & $\boldsymbol{\psi_0}$\\
    \hline
    Reissner-Nordström & 0 & -- & 0 & $7.500000 \times 10^{-1}$ & 0 & $-2.500000 \times 10^{-1}$ & $5.000000 \times 10^{-1}$\\
    \hline
    \multirow{3}{*}{Scalarized} & \multirow{3}{*}{0}  
      & 0 & $1.643270 \times 10^{-1}$ & $7.423045 \times 10^{-1}$ & $6.380654 \times 10^{-2}$ & $-2.517295 \times 10^{-1}$ & $4.992400 \times 10^{-1}$ \\ \cline{3-8}
    &  & 1 & $4.404460 \times 10^{-1}$ & $3.970001 \times 10^{-1}$ & $-4.039708 \times 10^{-1}$ & $-2.725778 \times 10^{-1}$ & $4.880798 \times 10^{-1}$ \\ \cline{3-8}
    &  & 1 & $1.658641 \times 10^{0}$ & $1.074043 \times 10^{-4}$ & $-2.651639 \times 10^{-1}$ & $-2.771950 \times 10^{-1}$ & $4.837580 \times 10^{-1}$ \\
    \hline
    \multirow{5}{*}{Scalarized} & \multirow{5}{*}{0.25} 
      & 0 & $1.643791 \times 10^{-1}$ & $7.423091 \times 10^{-1}$ & $6.387217 \times 10^{-2}$ & $-2.517308 \times 10^{-1}$ & $4.992396 \times 10^{-1}$ \\ \cline{3-8}
    &  & 1 & $4.403748 \times 10^{-1}$ & $3.995919 \times 10^{-1}$ & $-4.058366 \times 10^{-1}$ & $-2.726379 \times 10^{-1}$ & $4.880572 \times 10^{-1}$ \\ \cline{3-8}
    &  & 1 & $1.880295 \times 10^{0}$ & $5.859886 \times 10^{-5}$ & $-2.578252 \times 10^{-1}$ & $-2.777587 \times 10^{-1}$ & $4.833496 \times 10^{-1}$ \\ \cline{3-8}
    &  & 1 & $3.388940 \times 10^{0}$ & $1.991125 \times 10^{-6}$ & $-2.088096 \times 10^{-1}$ & $-2.771381 \times 10^{-1}$ & $4.834680 \times 10^{-1}$ \\ \cline{3-8}
    &  & 2 & $3.612684 \times 10^{0}$ & $2.087714 \times 10^{-6}$ & $7.144242 \times 10^{-1}$ & $-3.192613 \times 10^{-1}$ & $4.628959 \times 10^{-1}$ \\
    \hline
    \multirow{4}{*}{Scalarized} & \multirow{4}{*}{0.5} 
      & 0 & $1.644313 \times 10^{-1}$ & $7.423137 \times 10^{-1}$ & $6.393810 \times 10^{-2}$ & $-2.517322 \times 10^{-1}$ & $4.992392 \times 10^{-1}$ \\ \cline{3-8}
    &  & 1 & $4.403014 \times 10^{-1}$ & $4.022098 \times 10^{-1}$ & $-4.077352 \times 10^{-1}$ & $-2.727015 \times 10^{-1}$ & $4.880345 \times 10^{-1}$ \\ \cline{3-8}
    &  & 2 & $2.509742 \times 10^{0}$ & $3.442004 \times 10^{-5}$ & $7.884033 \times 10^{-1}$ & $-3.261751 \times 10^{-1}$ & $4.595228 \times 10^{-1}$ \\ \cline{3-8}
    &  & 3 & $2.574587 \times 10^{0}$ & $5.081207 \times 10^{-5}$ & $-2.159634 \times 10^{0}$ & $-4.670110 \times 10^{-1}$ & $4.048366 \times 10^{-1}$ \\
    \hline
    \multirow{4}{*}{Scalarized} & \multirow{4}{*}{1.5} 
      & 0 & $1.646413 \times 10^{-1}$ & $7.423323 \times 10^{-1}$ & $6.420470 \times 10^{-2}$ & $-2.517378 \times 10^{-1}$ & $4.992376 \times 10^{-1}$ \\ \cline{3-8}
    &  & 1 & $4.399836 \times 10^{-1}$ & $4.129494 \times 10^{-1}$ & $-4.156784 \times 10^{-1}$ & $-2.729636 \times 10^{-1}$ & $4.879416 \times 10^{-1}$ \\ \cline{3-8}
    &  & 2 & $1.369795 \times 10^{0}$ & $2.976045 \times 10^{-3}$ & $1.012055 \times 10^{0}$ & $-3.405495 \times 10^{-1}$ & $4.533391 \times 10^{-1}$ \\ \cline{3-8}
    &  & 3 & $1.481916 \times 10^{0}$ & $4.345425 \times 10^{-3}$ & $-3.066100 \times 10^{0}$ & $-5.293261 \times 10^{-1}$ & $3.870235 \times 10^{-1}$ \\
    \hline
    \multirow{3}{*}{Scalarized} & \multirow{3}{*}{5} 
      & 0 & $1.653907 \times 10^{-1}$ & $7.424008 \times 10^{-1}$ & $6.517607 \times 10^{-2}$ & $-2.517581 \times 10^{-1}$ & $4.992320 \times 10^{-1}$ \\ \cline{3-8}
    &  & 1 & $4.384612 \times 10^{-1}$ & $4.542177 \times 10^{-1}$ & $-4.489366 \times 10^{-1}$ & $-2.740087 \times 10^{-1}$ & $4.875927 \times 10^{-1}$ \\ \cline{3-8}
    &  & 2 & $7.779782 \times 10^{-1}$ & $1.587925 \times 10^{-1}$ & $1.740148 \times 10^{0}$ & $-3.693500 \times 10^{-1}$ & $4.457714 \times 10^{-1}$ \\
    \hline
    \multirow{2}{*}{Scalarized} & \multirow{2}{*}{10} 
      & 0 & $1.665029 \times 10^{-1}$ & $7.425086 \times 10^{-1}$ & $6.667797 \times 10^{-2}$ & $-2.517888 \times 10^{-1}$ & $4.992238 \times 10^{-1}$ \\ \cline{3-8}
    &  & 1 & $4.342744 \times 10^{-1}$ & $5.245427 \times 10^{-1}$ & $-5.208909 \times 10^{-1}$ & $-2.760121 \times 10^{-1}$ & $4.870304 \times 10^{-1}$ \\
    \hline
  \end{tabular}
  \caption{Values of the parameters $\gamma$, $n$, $\phi_H$, $a_H$, $\phi_0$, $b_1$ and $\psi_0$ for the RN and scalarized black hole solutions for $\eta=5$, $q=0.5$ and $\kappa=1$.}
  \label{parametersolutions}
\end{table*}

\begin{table*}
  \centering
  \scriptsize
  \begin{tabular}{|c|c|c|c|c|c|c|c|}
    \hline
    \textbf{Type of BH} & $\boldsymbol{\gamma}$ & \textbf{Nodes} & $\boldsymbol{\phi_H}$ & $\boldsymbol{a_H}$ & $\boldsymbol{\phi_0}$ & $\boldsymbol{b_1}$ & $\boldsymbol{\psi_0}$\\
    \hline
    Reissner-Nordström & 0 & -- & 0 & $7.500000 \times 10^{-1}$ & 0 & $-2.500000 \times 10^{-1}$ & $5.000000 \times 10^{-1}$\\
    \hline
    \multirow{4}{*}{Scalarized} & \multirow{4}{*}{0}  
      & 0 & $6.704430 \times 10^{-2}$ & $7.487109 \times 10^{-1}$ & $2.605226 \times 10^{-2}$ & $-2.502882 \times 10^{-1}$ & $4.998731 \times 10^{-1}$ \\ \cline{3-8}
    &  & 1 & $1.701177 \times 10^{-1}$ & $6.797663 \times 10^{-1}$ & $-1.730469 \times 10^{-1}$ & $-2.538307 \times 10^{-1}$ & $4.979711 \times 10^{-1}$ \\ \cline{3-8}
    &  & 2 & $3.517520 \times 10^{-1}$ & $1.883585 \times 10^{-1}$ & $4.909787 \times 10^{-1}$ & $-2.671836 \times 10^{-1}$ & $4.906749 \times 10^{-1}$ \\ \cline{3-8}
    &  & 2 & $6.890262 \times 10^{-1}$ & $2.918346 \times 10^{-3}$ & $3.964947 \times 10^{-1}$ & $-2.676134 \times 10^{-1}$ & $4.898022 \times 10^{-1}$ \\
    \hline
    \multirow{7}{*}{Scalarized} & \multirow{7}{*}{0.5} 
      & 0 & $6.704549 \times 10^{-2}$ & $7.487110 \times 10^{-1}$ & $2.605374 \times 10^{-2}$ & $-2.502882 \times 10^{-1}$ & $4.998731 \times 10^{-1}$ \\ \cline{3-8}
    &  & 1 & $1.701229 \times 10^{-1}$ & $6.797948 \times 10^{-1}$ & $-1.730841 \times 10^{-1}$ & $-2.538292 \times 10^{-1}$ & $4.979708 \times 10^{-1}$ \\ \cline{3-8}
    &  & 2 & $3.514915 \times 10^{-1}$ & $1.893383 \times 10^{-1}$ & $4.914183 \times 10^{-1}$ & $-2.671918 \times 10^{-1}$ & $4.906730 \times 10^{-1}$ \\ \cline{3-8}
    &  & 2 & $6.951567 \times 10^{-1}$ & $2.772346 \times 10^{-3}$ & $3.959627 \times 10^{-1}$ & $-2.676330 \times 10^{-1}$ & $4.897865 \times 10^{-1}$ \\ \cline{3-8}
    &  & 1 & $6.262757 \times 10^{0}$ & $5.610015 \times 10^{-10}$ & $-1.288726 \times 10^{-1}$ & $-2.572366 \times 10^{-1}$ & $4.955199 \times 10^{-1}$ \\ \cline{3-8}
    &  & 1 & $6.307691 \times 10^{0}$ & $5.652773 \times 10^{-10}$ & $-8.232402 \times 10^{-2}$ & $-2.570099 \times 10^{-1}$ & $4.955099 \times 10^{-1}$ \\ \cline{3-8}
    &  & 2 & $6.321481 \times 10^{0}$ & $5.810354 \times 10^{-10}$ & $2.585724 \times 10^{-1}$ & $-2.634100 \times 10^{-1}$ & $4.920063 \times 10^{-1}$ \\
    \hline
    \multirow{7}{*}{Scalarized} & \multirow{7}{*}{1.5} 
      & 0 & $6.704785 \times 10^{-2}$ & $7.487111 \times 10^{-1}$ & $2.605670 \times 10^{-2}$ & $-2.502882 \times 10^{-1}$ & $4.998731 \times 10^{-1}$ \\ \cline{3-8}
    &  & 1 & $1.701333 \times 10^{-1}$ & $6.798517 \times 10^{-1}$ & $-1.731583 \times 10^{-1}$ & $-2.538303 \times 10^{-1}$ & $4.979704 \times 10^{-1}$ \\ \cline{3-8}
    &  & 2 & $3.509765 \times 10^{-1}$ & $1.912926 \times 10^{-1}$ & $4.922963 \times 10^{-1}$ & $-2.672055 \times 10^{-1}$ & $4.906692 \times 10^{-1}$ \\ \cline{3-8}
    &  & 2 & $7.082807 \times 10^{-1}$ & $2.488080 \times 10^{-3}$ & $3.948268 \times 10^{-1}$ & $-2.676730 \times 10^{-1}$ & $4.897537 \times 10^{-1}$ \\ \cline{3-8}
    &  & 2 & $3.625208 \times 10^{0}$ & $4.741418 \times 10^{-8}$ & $2.789210 \times 10^{-1}$ & $-2.649871 \times 10^{-1}$ & $4.910447 \times 10^{-1}$ \\ \cline{3-8}
    &  & 3 & $3.645930 \times 10^{0}$ & $5.102971 \times 10^{-8}$ & $-7.012085 \times 10^{-1}$ & $-2.843759 \times 10^{-1}$ & $4.810074 \times 10^{-1}$ \\ \cline{3-8}
    &  & 4 & $3.650919 \times 10^{0}$ & $5.937503 \times 10^{-8}$ & $1.372727 \times 10^{0}$ & $-3.267005 \times 10^{-1}$ & $4.601055 \times 10^{-1}$ \\
    \hline
    \multirow{9}{*}{Scalarized} & \multirow{9}{*}{9} 
      & 0 & $6.706558 \times 10^{-2}$ & $7.487117 \times 10^{-1}$ & $2.607900 \times 10^{-2}$ & $-2.502884 \times 10^{-1}$ & $4.998731 \times 10^{-1}$ \\ \cline{3-8}
    &  & 1 & $1.702115 \times 10^{-1}$ & $6.802807 \times 10^{-1}$ & $-1.737201 \times 10^{-1}$ & $-2.538386 \times 10^{-1}$ & $4.979671 \times 10^{-1}$ \\ \cline{3-8}
    &  & 2 & $3.473405 \times 10^{-1}$ & $2.057713 \times 10^{-1}$ & $4.988672 \times 10^{-1}$ & $-2.673090 \times 10^{-1}$ & $4.906397 \times 10^{-1}$ \\ \cline{3-8}
    &  & 2 & $9.213339 \times 10^{-1}$ & $5.534701 \times 10^{-4}$ & $3.764800 \times 10^{-1}$ & $-2.680704 \times 10^{-1}$ & $4.893970 \times 10^{-1}$ \\ \cline{3-8}
    &  & 2 & $1.201825 \times 10^{0}$ & $1.176021 \times 10^{-4}$ & $3.537185 \times 10^{-1}$ & $-2.680455 \times 10^{-1}$ & $4.893102 \times 10^{-1}$ \\ \cline{3-8}
    &  & 3 & $1.445023 \times 10^{0}$ & $6.812403 \times 10^{-5}$ & $-8.235311 \times 10^{-1}$ & $-2.911957 \times 10^{-1}$ & $4.773418 \times 10^{-1}$ \\ \cline{3-8}
    &  & 4 & $1.482364 \times 10^{0}$ & $7.948659 \times 10^{-5}$ & $1.679312 \times 10^{0}$ & $-3.416378 \times 10^{-1}$ & $4.536186 \times 10^{-1}$ \\ \cline{3-8}
    &  & 5 & $1.489959 \times 10^{0}$ & $1.053493 \times 10^{-4}$ & $-3.075432 \times 10^{0}$ & $-4.377315 \times 10^{-1}$ & $4.150283 \times 10^{-1}$ \\ \cline{3-8}
    &  & 6 & $1.490683 \times 10^{0}$ & $1.581331 \times 10^{-4}$ & $5.412558 \times 10^{0}$ & $-6.116299 \times 10^{-1}$ & $3.615348 \times 10^{-1}$ \\
    \hline
    \multirow{6}{*}{Scalarized} & \multirow{6}{*}{15} 
      & 0 & $6.707979 \times 10^{-2}$ & $7.487122 \times 10^{-1}$ & $2.609690 \times 10^{-2}$ & $-2.502886 \times 10^{-1}$ & $4.998730 \times 10^{-1}$ \\ \cline{3-8}
    &  & 1 & $1.702737 \times 10^{-1}$ & $6.806269 \times 10^{-1}$ & $-1.741755 \times 10^{-1}$ & $-2.538453 \times 10^{-1}$ & $4.979645 \times 10^{-1}$ \\ \cline{3-8}
    &  & 2 & $3.446748 \times 10^{-1}$ & $2.171737 \times 10^{-1}$ & $5.041305 \times 10^{-1}$ & $-2.673934 \times 10^{-1}$ & $4.906150 \times 10^{-1}$ \\ \cline{3-8}
    &  & 3 & $1.092655 \times 10^{0}$ & $5.354462 \times 10^{-4}$ & $-8.911247 \times 10^{-1}$ & $-2.938847 \times 10^{-1}$ & $4.760090 \times 10^{-1}$ \\ \cline{3-8}
    &  & 4 & $1.143922 \times 10^{0}$ & $6.090433 \times 10^{-4}$ & $1.847102 \times 10^{0}$ & $-3.488130 \times 10^{-1}$ & $4.506537 \times 10^{-1}$ \\ \cline{3-8}
    &  & 5 & $1.153670 \times 10^{0}$ & $8.272398 \times 10^{-4}$ & $-3.530031 \times 10^{0}$ & $-4.568913 \times 10^{-1}$ & $4.090997 \times 10^{-1}$ \\
    \hline
    \multirow{5}{*}{Scalarized} & \multirow{5}{*}{30} 
      & 0 & $6.711540 \times 10^{-2}$ & $7.487135 \times 10^{-1}$ & $2.614187 \times 10^{-2}$ & $-2.502890 \times 10^{-1}$ & $4.998729 \times 10^{-1}$ \\ \cline{3-8}
    &  & 1 & $1.704282 \times 10^{-1}$ & $6.815037 \times 10^{-1}$ & $-1.753383 \times 10^{-1}$ & $-2.538622 \times 10^{-1}$ & $4.979578 \times 10^{-1}$ \\ \cline{3-8}
    &  & 2 & $3.387129 \times 10^{-1}$ & $2.452318 \times 10^{-1}$ & $5.174608 \times 10^{-1}$ & $-2.676109 \times 10^{-1}$ & $4.905498 \times 10^{-1}$ \\ \cline{3-8}
    &  & 3 & $7.421514 \times 10^{-1}$ & $8.287225 \times 10^{-3}$ & $-1.035708 \times 10^{0}$ & $-2.981407 \times 10^{-1}$ & $4.742154 \times 10^{-1}$ \\ \cline{3-8}
    &  & 4 & $8.063438 \times 10^{-1}$ & $8.866974 \times 10^{-3}$ & $2.236579 \times 10^{0}$ & $-3.631360 \times 10^{-1}$ & $4.453072 \times 10^{-1}$ \\
    \hline
  \end{tabular}
  \caption{Values of the parameters $\gamma$, $n$, $\phi_H$, $a_H$, $\phi_0$, $b_1$ and $\psi_0$ for the RN and scalarized black hole solutions for $\eta=30$, $q=0.5$ and $\kappa=1$.}
  \label{parametersolutions_eta30}
\end{table*}

\end{widetext}

\end{document}